\documentclass{aa}
\usepackage{graphics}
\usepackage{psfig}

\begin{document}

\newcommand{\al}    {\rm et al.}
\newcommand{\eg}    {\em e.g.}
\newcommand{\ie}    {\em i.e.}
\newcommand{\met}   {metallicity }
\newcommand{\kms}   {km s$^{-1}$}
\newcommand\beqa{\begin{eqnarray}}
\newcommand\eeqa{\end{eqnarray}}
\newcommand{\aaa} [2]{A\&A {\bf #1}, #2}
\newcommand{\aas} [2]{A\&A Suppl. {\bf #1}, #2}
\newcommand{\aj}  [2]{AJ {\bf #1}, #2}
\newcommand{\apj} [2]{Ap. J. {\bf #1}, #2}
\newcommand{\apjl}[2]{Ap. J. Letter {\bf #1}, #2}
\newcommand{\apjs}[2]{Ap. J. Suppl. {\bf #1}, #2}
\newcommand{\araa}[2]{A\&AR {\bf #1}, #2}
\newcommand{\pasp}[2]{PASP {\bf #1}, #2}
\newcommand{\mnras}[2]{MNRAS {\bf #1}, #2}

\title{A revised HRD for individual components of binary \\  
systems from BaSeL BVRI synthetic photometry.}
\subtitle{Influence of interstellar extinction and stellar rotation.} 
\author{E. Lastennet, J. Fernandes, and Th. Lejeune}
 
\institute{Observat\'orio Astron\'omico da Universidade de Coimbra, Santa Clara,
           P-3040 Coimbra, Portugal}

\offprints{E. Lastennet}

\date{Received 22 February 2002 / Accepted 20 March 2002}

\authorrunning{E. Lastennet {\al}}
\titlerunning{A revised HRD for individual components of binary systems with BaSeL}

\abstract{
Johnson BVRI photometric data for individual components of binary systems 
have been provided by ten Brummelaar {\al} (2000). 
This is essential because non interacting binaries can be considered as two 
single stars and therefore have to play a critical role in testing and calibrating 
single-star stellar evolution sets of isochrones and the implicit theory.
While they derived the effective temperature (T$_{\rm eff}$) from their estimated
spectral type, we infer metallicity-dependent T$_{\rm eff}$ from a minimizing 
method fitting the B$-$V, V$-$R and V$-$I colours. 
For this purpose, a grid of 621,600 flux distributions were computed from the Basel 
Stellar Library (BaSeL 2.2) of model-atmosphere spectra, and their theoretical colours
compared with the observed photometry. As a matter of fact, the BaSeL colours show a 
very good agreement with the BVRI metallicity-dependent empirical calibrations of 
Alonso {\al} (1996), temperatures being different by 3$\pm$3\% in the range 
4000-8000 K for dwarf stars. 
Before deriving the metallicity-dependent T$_{\rm eff}$ from the BaSeL models, 
we paid particular attention to the influence of reddening and stellar rotation.  
We inferred the reddening from two different methods: (i) the MExcessNg code v1.1 
(M\'endez \& van Altena 1998) and (ii) neutral hydrogen column density data. 
A comparison of both methods shows a good agreement for the sample which is located 
inside a local sphere of $\sim$500 pc, but we point out a few directions where 
the MExcess model overestimates the E(B$-$V) colour excess. 
Influence of stellar rotation on the BVRI colours can be neglected except for 
5 stars with large $v \sin i$, the maximum effect on temperature being less than 5\%. 
Our final determinations provide effective temperature estimates for each 
component. They are in good agreement with previous spectroscopic determinations 
available for a few primary components, and with ten Brummelaar {\al} below 
$\sim$10,000 K. Nevertheless, we obtain an increasing disagreement with their 
temperatures beyond 10,000 K. 
Finally, we provide a revised Hertzsprung-Russell diagram (HRD) for 
the systems with the more accurately determined temperatures.  
}

\maketitle

\section{Introduction}

The knowledge of the fundamental parameters of binary system members is essential because 
the calibration of binary stars on the HR diagram can be used to determine {\eg} the 
helium abundance, helium-to-heavier-elements ratio, age and mixing length parameter for 
stars other than the Sun (see {\eg} Fernandes {\al} 1998, Lastennet {\al} 1999b, 
Lebreton {\al} 2001). 
Recently, ten Brummelaar {\al} (2000, hereafter tB00) have provided Johnson BVRI 
photometric data for individual components of visual binary systems. This is a rare  
opportunity to derive their individual effective temperature from photometric 
calibrations, and hence to place the stars on a HR diagram.  
In this paper, we use the BaSeL models in Johnson BVRI photometry to derive the 
metallicity-dependent temperature of 56 stars, all binary systems members. 
Due to their large angular separation (see Table 1 of tB00), these systems should not 
be in contact so they can be assumed as single stars and provide possible candidates 
for future comparisons with evolutionary models. 
Most of these stars already have a T$_{\rm eff}$ determination derived in two steps 
by tB00: first a spectral type is estimated from each 
colour (B$-$V, V$-$R and V$-$I) with the Johnson's (1966) calibration tables, 
then from each spectral type a T$_{\rm eff}$ is derived from Landolt-B\"ornstein (1980). 
Such a method is a good first order approximation, but possible errors 
can accumulate faster with the addition of two calibration methods (colour-spectral 
type plus spectral type-T$_{\rm eff}$) so that their assigned uncertainties 
might be too optimistic. 
Moreover, we intend to improve the study of ten Brummelaar {\al} (2000) by 
taking into acount the influence of interstellar extinction and stellar rotation. 
For these reasons, we present new T$_{\rm eff}$ values, homogeneously determined 
with the Basel Stellar Library (BaSeL),   
a library of theoretical spectra corrected to provide synthetic colours consistent 
with empirical colour-temperature calibrations at all wavelengths from the near-UV 
to the far-IR and covering a large range of fundamental parameters 
(see Lejeune {\al} 1998 and references therein for a complete description).  
The BaSeL models have already been used to determine fundamental parameters with 
success, both in broad or/and medium-band photometry (see Lejeune \& Buser 1996, 
Lejeune 1997, Lastennet {\al} 1999a, 2001, 2002).  
In this paper we intend to use them in the Johnson photometric system. 
Therefore, we strongly stress that in the remainder of this paper {\it BVRI will 
stand for Johnson photometry, not the BVRI$_C$ Johnson-Cousins photometry}.
In order to assess the quality of our BaSeL-derived T$_{\rm eff}$s, we obtained 
very good agreement with the Alonso {\al} (1996) empirical calibrations. 
Furthermore, some of the individual components have spectroscopic determinations 
providing a stringent test to the T$_{\rm eff}$s derived by tB00 and the BaSeL models. 
In addition, the Marsakov \& Shevelev (1995) catalogue provides further comparisons. \\
The paper is organized as follows: Sect. 2 deals with the description of our working 
sample and the method used to derive metallicity-dependent T$_{\rm eff}$  
with the BaSeL library along with other sources of determinations.     
Sect. 3 presents the extinction issue and the influence of stellar rotation, and 
Sect. 4 is devoted to the presentation and the discussion of the results and the 
revised HR diagram.   
Finally, Sect. 6 draws our general conclusions. 

\section{Relevant data of the working sample and description of the method} 

\subsection{Working sample and relevant data}
We have selected binary stars with at least 2 colours from the list of ten Brummelaar 
{\al} (2000). This selection gives 28 systems ({\ie} 56 individual 
components).  
Identifications (arbitrary ID number and HD numbers), galactic coordinates and parallaxes 
($l$,$b$ and $\pi$ from 
SIMBAD\footnote{The SIMBAD parallaxes are from the Hipparcos catalogue 
(ESA, 1997, see also Perryman {\al}, 1997), except for the systems [12], [14] and [16].}, 
as well as two determinations of the E(B$-$V) colour 
excess (discussed in \S\ref{s:extinction})  
along with the adopted E(B$-$V) values,
and projected rotational velocities are presented in Table 1. 

\begin{table*}[htb] 
\caption[]{Working sample of visual binaries: 
cross-identifications (ID is an arbitrary running number), galactic coordinates and parallaxes 
($l$,$b$ and $\pi$ from SIMBAD). The E(B$-$V) colour excess derived from 1) the MExcess code 
of M\'endez \& van Altena (1998) from $l$, $b$ and $\pi$ (errors on $\pi$ are propagated on E(B$-$V)) 
and 2) N$_{HI}$ data 
(if there is no measurement, the constraint is estimated from the nearest sources) 
is given along with the adopted colour excess in B$-$V (see text for E(V$-$R) and 
E(V$-$I)). Rotational velocities $v \sin i$ are given in the two last columns (mean value from SIMBAD and 
from the Glebocki \& Stawikowski catalogue, hereafter GS00).} 
\label{tab:data}
\begin{flushleft}
\begin{center}  
\begin{tabular}{rrrrrrrrrrrr}
\hline 
\noalign{\smallskip} 
 ID            & HD  & l      & b      & $\pi$  & $\sigma_{\pi}$/$\pi$ & \multicolumn{3}{c}{E(B$-$V)} &  \multicolumn{2}{c}{$v \sin i$} \\
               &     & [$^o$] & [$^o$] & [mas]  &   [10$^{-2}$]        &  \multicolumn{3}{c}{[mag]}   &  \multicolumn{2}{c}{[km s$^{-1}$]} \\                    
               &     &        &        &        &                      & MExcess    &  N$_{HI}$ data & Adopted & SIMBAD             &  GS00   \\                    
\noalign{\smallskip}
\hline \noalign{\smallskip}
1  &  224930 & 109.61 &$-$34.51 & 80.63$\pm$3.03 &  3.8  & 0.001$\pm$0.000            & 0.000       & 0.000  &3  &   1.8$\pm$0.6 \\
2  &  2772   & 120.05 &$-$8.24  &  9.20$\pm$1.06 & 11.5  & 0.418$^{+0.044}_{-0.043}$  & 0.007-0.156 & 0.045  &   &                   \\
3  &  13594  & 137.10 &$-$13.10 & 24.07$\pm$0.96 &  4.0  & 0.024$\pm$0.001            & $\leq$0.001 & 0.001  &   &  27.2$\pm$0.8 \\   
4  &  26722  & 183.60 &$-$28.94 &  8.78$\pm$1.39 & 15.8  & 0.059$^{+0.009}_{-0.007}$  & 0.000-0.142 & 0.030  &   &   5.1$\pm$1 \\
5  &  27820  & 185.11 &$-$26.88 &  8.23$\pm$0.94 & 11.4  & 0.076$^{+0.008}_{-0.006}$  & 0.001-0.265 & 0.160  &70 &  70$\pm$8 \\
6  &  28485  & 180.78 &$-$21.88 & 22.93$\pm$1.25 &  5.5  & 0.032$\pm$0.002            & $\leq$0.004 & 0.000  &150 & 165$\pm$10 \\
7  &  30810  & 187.99 &$-$20.48 & 20.15$\pm$1.14 &  5.7  & 0.026$\pm$0.002            & $\leq$0.004 & 0.002  & 3 &   5 \\
8  &  37468  & 206.82 &$-$17.34 &  2.84$\pm$0.91 & 32.0  & 0.205$^{+0.048}_{-0.035}$  & 0.059$^{+0.014}_{-0.011}$$^{(*)}$ & 0.048$^{(a)}$  &   &  86 \\
9  &  37711  & 190.09 &$-$7.31  &  4.36$\pm$1.23 & 28.2  & 0.288$^{+0.058}_{-0.020}$  & 0.007-0.326 & 0.007  &   &  95 \\
10 &  50522  & 157.77 &$+$23.60 & 19.14$\pm$0.76 &  4.0  & 0.010$^{+0.000}_{-0.001}$  & $\leq$0.001 & 0.000  & 13 &   3.2$\pm$1.1 \\
11 &  76943  & 179.80 &$+$41.18 & 60.86$\pm$1.30 &  2.1  & 0.001$\pm$0.000            & $\leq$0.001 & 0.000  & 25 &  23.6$\pm$1 \\
12 & 98231/0 & 195.11 &$+$69.25 & 130.$\pm$2.$^{(\dag)}$ & 1.5  &  0.000$\pm$0.000    & 0.000       & 0.000  & 10 &   2.8$\pm$0.7 \\ 
13 &  114330 & 311.42 &$+$57.03 &  7.86$\pm$1.11 & 14.1  &  0.009$\pm$0.000           & $\leq$0.055 & 0.055  & 10 & $<$10          \\
14 &  114378 & 327.93 &$+$79.49 & 51.$\pm$4      &  7.8  &  0.000$\pm$0.000           & 0.000$^{(*)}$ & 0.000  &  24 &  21.3$\pm$1   \\                                        
15 &  133640 & 80.37  &$+$57.07 & 78.39$\pm$1.03 & 1.31  &  0.001$\pm$0.000           & $\leq$0.001 & 0.000  & 15 &   1.9$\pm$0.5  \\  
16 &  137107 & 47.54  &$+$56.73 & 61.$\pm$4      &  6.6  &  0.000$\pm$0.000           & $\leq$0.004 & 0.000  &    &   2.8$\pm$0.7 \\ 
17 &  140436 & 41.74  &$+$51.92 & 22.48$\pm$0.67 &  3.0  &  0.010$^{+0.001}_{-0.000}$ & $\leq$0.001 & 0.000  & 100 & 100$\pm$8    \\ 
18 &  148857 & 17.12  &$+$31.84 & 19.63$\pm$1.34 &  6.8  &  0.013$\pm$0.001           & $\leq$0.005 & 0.000 & 142 & 125$\pm$8  \\ 
19 &  155125 & 6.72   &$+$14.01 & 38.77$\pm$0.86 &  2.2  &  0.024$^{+0.000}_{-0.001}$ & $\leq$0.004 & 0.000  & 14  & 15$\pm$8   \\  
20 &  188405 & 34.15  &$-$17.26 & 11.67$\pm$1.18 & 10.1  &  0.023$^{+0.003}_{-0.002}$ & $\leq$0.076 & 0.076  &     &           \\ 
21 &  190429 & 72.59  &$+$2.61  &  0.03$\pm$1.02 & 3400. &  0.465$^{+0.000}_{-0.190}$ & $\geq$0.057 & 0.057$^{(b)}$  & 170 & 105-135\\
22 &  193322 & 78.10  &$+$2.78  &  2.10$\pm$0.61 &  2.9  &  0.162$^{+0.070}_{-0.038}$ & 0.253$^{+0.121}_{-0.082}$$^{(*)}$ & 0.205$^{(c)}$  & 200 &  67-86  \\                                                            
23 &  196524 & 58.88  &$-$15.65 & 33.49$\pm$0.88 &  2.6  &  0.006$^{+0.001}_{-0.000}$ & $\leq$0.018 & 0.000$^{(d)}$ & 55 &  40$\pm$0.4   \\ 
24 &  200499 & 28.05  &$-$37.86 & 20.64$\pm$1.47 &  7.1  &  0.010$\pm$0.001           & $\leq$0.005 & 0.000  & 53 &  53$\pm$8     \\  
25 &  202275 & 60.49  &$-$25.66 & 54.11$\pm$0.85 &  1.6  &  0.004$\pm$0.000           & $\leq$0.018 & 0.010  & 13 &  5$\pm$1.5    \\ 
26 &  202444 & 82.85  &$-$7.43  & 47.80$\pm$0.61 &  1.3  &  0.575$^{+0.001}_{-0.000}$ & 0.000$^{(*)}$ &  0.000$^{(e)}$ & 91 & 98$\pm$10  \\  
27 &  202908 & 62.55  &$-$25.51 & 19.79$\pm$1.18 &  6.0  &  0.010$\pm$0.000           & $\leq$0.006 & 0.006  &    &  8$\pm$1    \\ 
28 &  213235 & 70.22  &$-$43.50 & 18.93$\pm$1.23 &  6.5  &  0.015$\pm$0.001           & $\leq$0.004 & 0.004  & 70 & 65       \\
\noalign{\smallskip}\hline        
\end{tabular}
\end{center}
\newpage
$^{(\dag)}$ The result on E(B$-$V) is unchanged with the more 
recent parallax of S\"oderhjelm (1999): $\pi$$=$119.7$\pm$0.8 mas.
$^{(*)}$ This star has a $N_{HI}$ measurement.
(a) Adopted value for the system but $\chi$$^2$ experiments with the BaSeL models would 
suggest E(B$-$V)$\sim$0.005 which is ruled out by the N$_{HI}$ data.  
(b) Another minimum exists at E(B$-$V)$=$0.10 for the secondary. 
(c) The BaSeL models suggest E(B$-$V)$=$0 for the secondary, but this is ruled out by 
the N$_{HI}$ data. 
(d) Adopted value but $\chi$$^2$ experiments with the BaSeL models would 
suggest E(B$-$V)$\sim$0.02 (upper limit derived from N$_{HI}$ data) for the secondary component.
(e) $\chi$$^2$ experiments with the BaSeL models suggest E(B$-$V)$\sim$0.05 for the 
primary component, which is ruled out by the N$_{HI}$ data.  
\end{flushleft}
\end{table*}

A remark has to be made at this point about the sensitivity of the B$-$V, V$-$R and 
V$-$I data used in this paper to the T$_{\rm eff}$, [Fe/H] and log g stellar parameters. 
While these colours are intrinsically sensitive to the T$_{\rm eff}$,  
V$-$R and V$-$I are known to be insensitive to surface gravity and not very sensitive to 
metallicity as shown by Buser \& Kurucz (1992) in BVRI$_C$ photometry, 
so no strong constraint can be expected on these two atmospheric parameters. 
However, we present metallicity-dependent 
T$_{\rm eff}$ determinations, which means that we can directly predict a temperature 
range for any given metallicity. This will be of particular interest when a spectroscopic 
[Fe/H] determination is available. \\
In the next subsections, we present the method used to derive T$_{\rm eff}$ from the 
BaSeL models along with earlier determinations. 
\begin{figure}[htb]
\centerline{\psfig{file=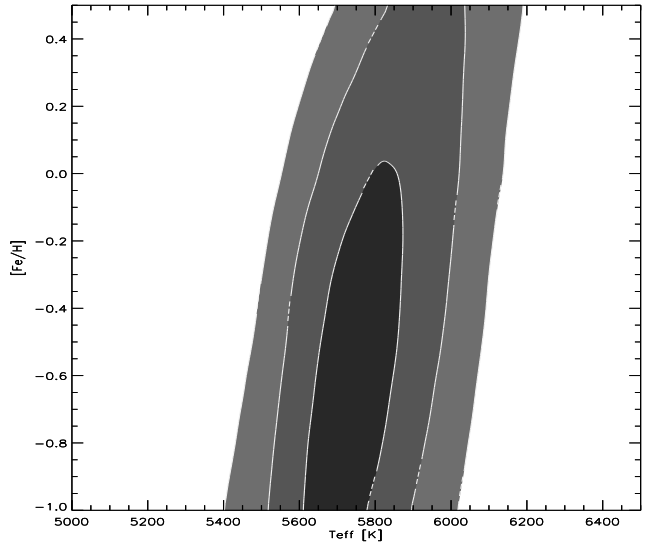,width=9.truecm,height=7.5truecm,rheight=7.9truecm}}
\caption{Example of contour solutions (1-, 2- and 3- $\sigma$, 1$\sigma$ being in black) 
in the (T$_{\rm eff}$, [Fe/H]) plane by fitting the dereddened 
observed (B$-$V), (V$-$R) and (V$-$I) colours of the primary component of $\xi$ UMa 
[12] with the BaSeL library.   
}
\label{f:contours}
\end{figure}

\subsection{Best fitting method applied to the BaSeL models} 
For the present work, more than 621,600 models have been computed by interpolation 
with the Interactive BaSeL Server (Lejeune, 2002), each model giving synthetic 
photometry in the Johnson system for a set of 
(T$_{\rm eff}$, [Fe/H], log g). 
In order to fit the observed colours of the sample stars, we have computed a fine 
grid in the (T$_{\rm eff}$, [Fe/H], log g) parameter space. 
The grid explored is defined in this way: 3000 $\leq$ T$_{\rm eff}$ $\leq$ 40,000 K in 
20 K steps, $-$1 $\leq$ [Fe/H] $\leq$ 0.5 in 0.1 steps, 
and 3 $\leq$ log g $\leq$ 5 in 0.1 steps.

In order to derive simultaneously the effective temperature (T$_{\rm eff}$), the \met 
([Fe/H]), and the surface gravity (log g) of each star, we minimize the $\chi^2$-functional 
defined as: \\
\beqa 
\chi^2 & = &
\sum_{i=1}^{3} \left[ \left(\frac{\rm col(i)_{\rm BaSeL} - col(i)_{\rm Obs.}}
{\sigma(\rm col(i)_{\rm Obs.})}\right)^2 \right]   \nonumber \\
\eeqa
where col(i)$_{\rm Obs.}$ and $\sigma$(col(i)$_{\rm Obs.}$) are the observed 
values (B$-$V, V$-$R, V$-$I) and their uncertainties from tB00, 
and col(i)$_{\rm BaSeL}$ is obtained from the synthetic
computations of the BaSeL models. 
A similar method has already been developed and used by Lastennet {\al} 
(1996) for CMD diagrams (Lastennet {\al} 1999b)
and for COROT potential targets (Lastennet {\al} 2001). 
Basically, small $\chi^2$ values are signatures of good fits. 
A $\chi^2$-grid is formed in the (T$_{\rm eff}$, [Fe/H], log g) parameter space. 
Once the central minimum value $\chi^{2}_{\rm min}$ is found, we compute the surfaces  
corresponding to 1$\sigma$, 2$\sigma$, and 3$\sigma$ confidence levels, 
as shown on the example of Fig.~\ref{f:contours}. 
The uncertainties are therefore as realistic as possible because directly dependent 
on the uncertainty of the photometric data (see Eq. 1). 

\subsubsection{Comparison with the Alonso {\al} (1996) empirical calibrations in BVRI 
photometry}
Alonso {\al} (1996) provided BVRI metallicity-dependent empirical calibrations of the 
effective temperature, using the Infrared Flux Method Method on a large sample of 
stars (410 for B$-$V and 163 for VRI colours). 
The BaSeL models (version 2.2) are based on model-atmosphere spectra calibrated on the 
photometric fluxes, using for this empirical colour-temperature relations at solar 
metallicity and semi-empirical ones for non-solar metallicities (see Lejeune {\al} 1997, 
1998 for details about the calibration procedure). 
While it is beyond the scope of this paper to give a detailed analysis comparing the BaSeL 
library with empirical calibrations, it is useful to point out some comparisons 
with the comprehensive study of Alonso {\al} (1996). 
For dwarf stars, and considering only the colours 
(B$-$V, V$-$R and V$-$I), metallicities (ranging from $-$1.0 to 0.5) 
and temperatures hotter than 4000 K relevant for the present study, 
the difference between Alonso {\al} and BaSeL is about 3\% (and within 9\% in the worst case, 
{\ie} for V$-$R close to 4000 K and [Fe/H]$=$0.5). More detailed comparisons are  
shown in Lejeune (2002), but this excellent agreement stongly confirms the BaSeL 
prediction capabilities in the Johnson 
photometric system and fully justify their use for many astrophysical applications. \\ 
Moreover, the Alonso {\al} (1996) calibrations do not allow to cover the range of 
temperatures we are interested in this paper, because their relationships are only valid 
below $\sim$8000 K in B$-$V, and even lower in V$-$R ($\sim$7600 K) and V$-$I ($\sim$6800 K).  
Since these upper limits can be even lower according 
to [Fe/H], the BaSeL library appears to be ideal and accurate enough for the purpose of 
the present work. 

\subsection{Other determinations}
As already mentioned in the Introduction, tB00 determined the T$_{\rm eff}$ for 
most of the sample. Their results are reported in Tab.~\ref{tab:finalteff} and will 
be discussed in \S\ref{section:results}. \\
To be as complete as possible, we looked for other determinations available in the 
literature and the SIMBAD database. 
Marsakov \& Shevelev (1995) (hereafter MS95) have computed effective temperatures 
and surface gravities using Moon's (1985) method, which is also based on the 
interpolation of the grids presented in Moon \& Dworetsky (1985). 
According to Moon (1985), the standard deviation of the derived 
parameters are T$_{\rm eff}$$=$$\pm$ 100 K.
All the MS95 T$_{\rm eff}$s of our sample are given in Tab.~\ref{tab:finalteff}. \\
One of the most comprehensive compilation for our purpose is the 2001 Edition of the 
Cayrel de Strobel {\al} catalogue, which includes [Fe/H] determinations and atmospheric 
parameters ($T_{\rm eff}$, log g) obtained from high-resolution spectroscopic 
observations and detailed analyses, most of them carried out with the help of model 
atmospheres. Since this new version is restricted to intermediate and low-mass stars 
(F, G, and K stars), we also checked the previous issue of the compilation (Cayrel de 
Strobel {\al} 1997). Some of our sample stars are included in these catalogues and 
the T$_{\rm eff}$s are also reported in Tab.~\ref{tab:finalteff}. 

\section{Influence of interstellar extinction and stellar rotation}
%
\subsection{Reddening}
\label{s:extinction}
\begin{figure}[htb]
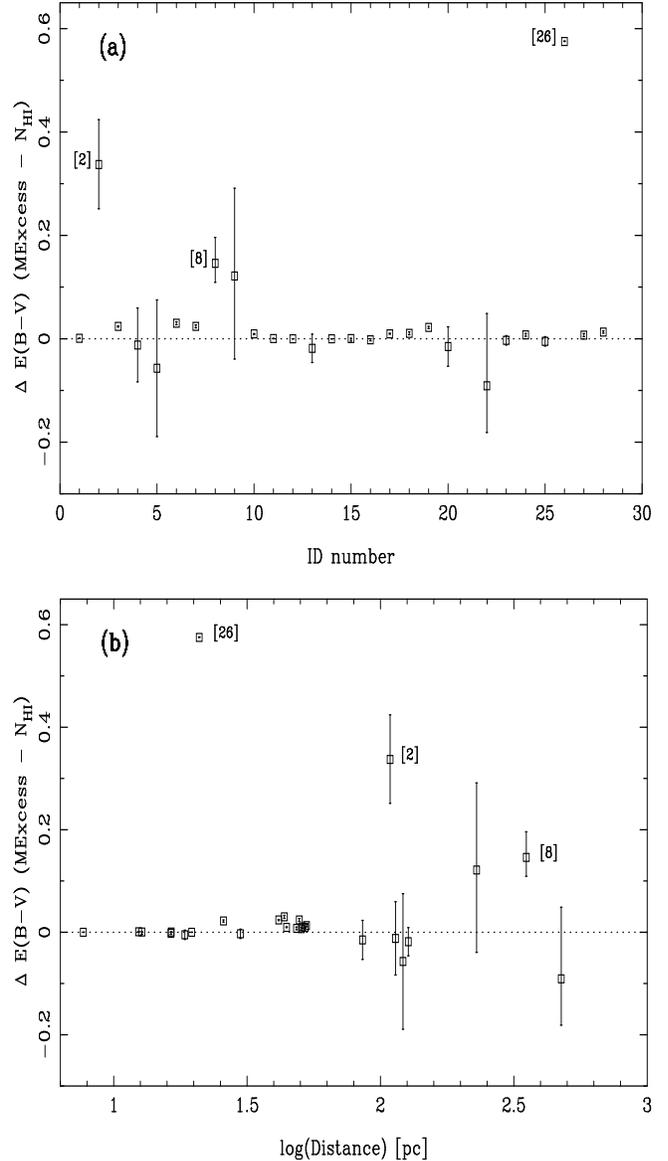

\centerline{\psfig{file=ebv1.ps,width=8.5truecm,height=7.5truecm,rheight=7.9truecm,angle=-90.}}
\centerline{\psfig{file=ebv2.ps,width=8.5truecm,height=7.5truecm,angle=-90.}}
\caption{
(a) Comparison between the E(B$-$V) values derived from the MExcess model 
and derived from N$_{HI}$ data for the objects (ID numbers) listed in Tab.~\ref{tab:data}. 
HD 190429 ([21]) is not shown because we only have a lower limit from N$_{HI}$ data and 
because its parallax is highly uncertain (see Tab.~\ref{tab:data}). 
The MExcess model overestimates the reddening for the binaries [2], [8] 
and [26], otherwise there is a good agreement. Panel (b) shows that there 
is no systematics with the distance and that both determinations present an expected 
very small reddening in the close solar neighbourhood ($\sim$60 pc).      
}
\label{f:ebv}
\end{figure}

Because estimates of the interstellar extinction are required for any photometric 
calibration method, we pay particular attention to the reddening before deriving 
any result with the BaSeL library. 
For each star we took the galactic coordinates ($l$ and $b$) from
SIMBAD, as well as the parallaxe from which we derived the distance $d$.
Fixing ($l$,$b$,$d$), then we used the MExcessNg code v1.1 (M\'endez, van Altena
and Ng) based on the code developed by M\'endez \& van Altena (1998),
to derive the E(B$-$V) colour excess. The results we derived are listed in 
Tab.~\ref{tab:data} (MExcess column). \\
Before adopting the colour excess values provided by this code, 
we performed some $\chi$$^2$ experiments with the BaSeL models to check 
basic consistency. While good agreement was found in many cases, we found 
some unexpected results.  
For instance, the high excess colour found with the above mentioned code 
(E(B$-$V)$\sim$0.418) for the system HD 2772 (d$\sim$110 pc) 
is in strong disagreement with $\chi$$^2$-computations ($\chi$$^2$$\sim$76.). 
A good fit of the HD 2772 VRI colours is obtained with a much smaller E(B$-$V) 
value: $\chi$$^2$$\sim$0.3 for E(B$-$V) ranging from 0. to 0.05. 
Such an example of possible source of errors encouraged us to use another 
method. \\
%
\begin{table*}[htb] 
\caption[]{Influence of stellar rotation on the (B$-$V) colour index: minimum (maximum) 
effect is computed from Eq.~\ref{e:bv} with h$=$0.5 (2) and the lower (upper) observational 
limit of $v \sin i$ from GS00 catalogue (see Tab.~\ref{tab:data}).}
\label{tab:rotation}
\begin{flushleft}
\begin{center}  
\begin{tabular}{rcccccc}
\hline 
\noalign{\smallskip}
 ID$^{(\dag)}$ &  $v \sin i$  & $\Delta$(B$-$V) & 
\multicolumn{2}{c}{ $\Delta$(T$_{\rm eff}$)} & \multicolumn{2}{c}{ \underline{$\Delta$(T$_{\rm eff}$)$_{\rm max}$} }  \\
               &                  & min / max          &    &   &  \multicolumn{2}{c}{T$_{\rm eff}$} \\ 
               &   [km s$^{-1}$]  &   [mag.]  & \multicolumn{2}{c}{[K]} & \multicolumn{2}{c}{[\%]} \\
               &                  &            & p  & s  & p  & s \\
\noalign{\smallskip}
\hline \noalign{\smallskip}
6  & 165$\pm$10   & 0.012 / 0.061 &  \multicolumn{2}{c}{150$\pm$150$^{(\ddag)}$ } & \multicolumn{2}{c}{4}  \\
17 & 100$\pm$8    & 0.004 / 0.023 &  260$\pm$180 & 150$\pm$110         & 4 & 3  \\
18 & 125$\pm$8    & 0.007 / 0.035 &  \multicolumn{2}{c}{320$\pm$220$^{(\ddag)}$ } & \multicolumn{2}{c}{5}   \\
21 & 120$\pm$15   & 0.006 / 0.036 &  260$\pm$260 & 220$\pm$150                    & 5 & 4 \\
26 &  98$\pm$10   & 0.004 / 0.023 &  80$\pm$80   & 50$\pm$50                      & 2.5 & 1.5   \\
\noalign{\smallskip}\hline
\end{tabular}
\end{center}
$^{(\dag)}$ Running number as in Tab.~\ref{tab:data}. 
$^{(\ddag)}$ Approximation from the combined photometry (see text). \\
\end{flushleft}
\end{table*}
%
A better approach would be to use colour excess determinations for each star. 
This can be done from values of neutral hydrogen column density $N_{HI}$. 
For this we used the ISM Hydrogen Column Density Search 
Tool\footnote{http://archive.stsci.edu/euve/ism/ismform.html} using the compilation 
of data ({\eg} from satellite missions like ROSAT and EUVE) by Fruscione {\al} 
(1994) plus additional IUE measurements from Diplas \& Savage (1994). 
Giving the position already used with the MExcess code ($l$,$b$,$d$), 
this tool provides $N_{HI}$ measurements for the ten sources nearest to the point in space 
selected and, even better, a determination for 4 stars of our working 
sample ([8], [14], [22] and [26]). \\
With the observational estimates (or at least observational constraints) listed 
in Table 1, we derived 
E(B$-$V) adopting the following relation between E(B$-$V) and $N_{HI}$ (cm$^{-2}$): 
E(B$-$V)$=$1.75 10$^{-22}$$\times$log($N_{HI}$) 
(see Rucinski \& Duerbeck, 1997 and references therein with coefficient ranging 
from 1.7 to 1.8 10$^{-22}$). \\
Finally, to derive the colour excess in V$-$R and V$-$I, we used the following 
relations: 
E(V$-$I)$=$ 1.527$\times$E(B$-$V), 
E(V$-$R)$=$ 0.725$\times$E(B$-$V) according to Wegner (1994). \\
A comparison of both methods to infer the E(B$-$V) colour excess shows a quite 
good agreement (see Fig.~\ref{f:ebv} (a)) except for the binaries [2] (HD 2772, the 
example discussed before), [8] and [26] for which the MExcess model overestimates the 
colour excess. 
Moreover, there is no systematics between both methods with the distance (see 
Fig.~\ref{f:ebv} (b)).   
This gives some weight to the validity of the MExcess model, even if we stress 
that our working sample is small and that we have detected 3 anomalies. 
It is also interesting to note that both determinations present - as expected - 
a very small reddening in the close solar neighbourhood (within $\sim$60 pc on panel (b)). \\

\subsection{Rotation}
\label{s:rotation}
Stellar rotation is known to have an influence on photometric data (see {\eg} Maeder 1971, 
Collins \& Sonneborn 1977, Collins \& Smith 1985): basically a rotating star is similar 
to a non-rotating one with smaller effective temperature.  
Therefore its possible influence on the BVRI Johnson colours have to be studied. 
This can be neglected for our sample except when the $v \sin i$ is large. If only 
considering $v \sin i$ larger than 100 km s$^{-1}$ in Tab.~\ref{tab:data}, the effect of rotation has  
to be determined for only 5 binaries : [6], [17], [18], [21] and [26].   

In order to assess the {\it minimum} effect induced by rotation, we assume a 
uniform rotation (see {\eg} Collins, 1966) keeping in mind that a differential 
rotation (Collins \& Smith, 1985) produces a larger effect. 
To first order\footnote{This approximation was applied in simulations of open clusters CMDs 
by Lastennet \& Valls-Gabaud (1996) and Lastennet (1998).}, we assume that the B$-$V colour 
difference between rotating and non-rotating copartners is: \\
\begin{equation}
\indent \Delta (B-V)\; \sim \; 10^{-2} \; h \; 
\left(\frac{v \sin i}{100 \; {\rm km s^{-1}}}\right)^2 
\label{e:bv}
\end{equation}
%
where $v \sin i$ is the projected rotational velocity of a star, $v$ is the equatorial 
velocity and {\it i} is the inclination of the rotation axis to the line--of--sight.
Typical constant values in Eq.~\ref{e:bv} is $ h \approx 2^{\pm 1}$ 
(see Maeder \& Peytremann 1970, 1972)    
and may change with spectral type, age and chemical composition (Zorec, 1992).\\
%
\begin{figure}[htb]
\centerline{
\psfig{file=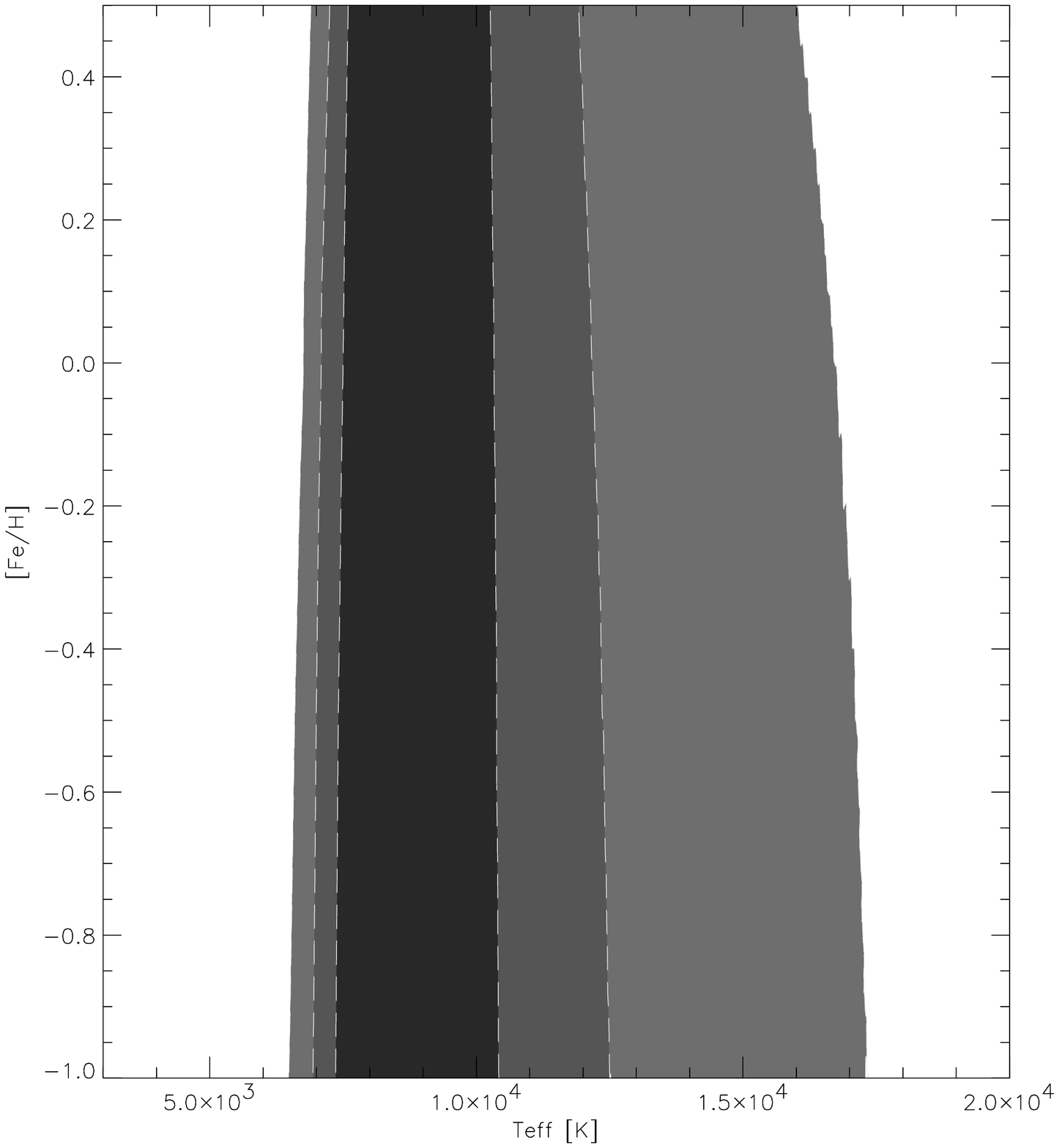,width=9.truecm,height=7.5truecm}
}
\centerline{
\psfig{file=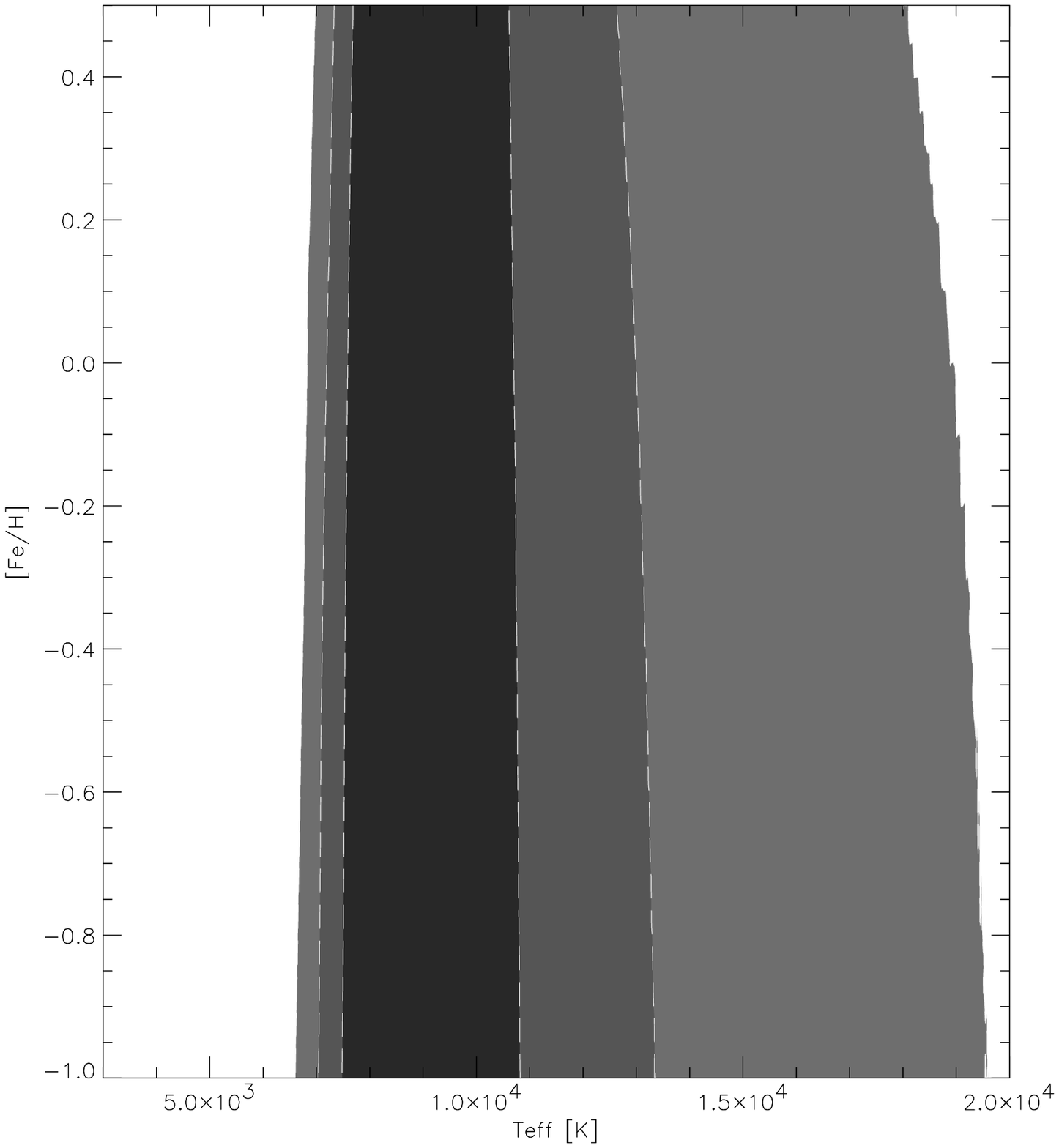,width=9.truecm,height=7.5truecm}
}
\caption{Effect of stellar rotation on the secondary component of HD 140436 ([17]). 
Contour solutions (1-, 2- and 3- $\sigma$, 1$\sigma$ being in black) 
are shown in the (T$_{\rm eff}$, [Fe/H]) plane by fitting the observed (B$-$V) 
colour: before correcting for rotation (upper panel) and after correction (lower panel). 
While the shape of the contour solutions is only slightly modified, 
there is a shift in the sense the T$_{\rm eff}$ increases when the influence 
of rotation on the B$-$V colour is taken into account ($\Delta$T$_{\rm eff}$$=$260 K).       
}
\label{f:rotation}
\end{figure}
The influence of rotation on the (B$-$V) colours of the 5 binaries that might be affected 
is summarized in Tab.~\ref{tab:rotation} where the shift to be applied is $\Delta$(B$-$V). 
Since the BV photometry is not available for each component of HD 28485 [6] and HD 148857 [18], 
we only use the combined photometry available in SIMBAD (B$-$V$=$0.32 [6] and 
B$-$V$=$0.01 [18]) as a rough approximation. 
In summary, the corrected B$-$V colours are given by (B$-$V)$_{\rm corrected}$$=$ 
(B$-$V) $-$ $\Delta$(B$-$V). Using (B$-$V)$_{\rm corrected}$, the BaSeL 
models were run again to derive the corrected T$_{\rm eff}$, hence providing 
the $\Delta$T$_{\rm eff}$ $=$ T$_{\rm eff,}$$_{\rm corrected}$ $-$ T$_{\rm eff}$. 
To illustrate this effect, Fig.~\ref{f:rotation} shows the BaSeL solutions fitting 
the B$-$V colours of the secondary component of HD 140436 ([17]): before correcting for rotation 
(upper panel) and after correction (lower panel), assuming the maximum influence reported in 
Tab.~\ref{tab:rotation} ($\Delta$(B$-$V)$=$0.023).  
While the shape of the contour solutions is only slightly modified, the T$_{\rm eff}$ increases 
when the influence of rotation on the B$-$V colour is taken into account: we measure a  
$\Delta$T$_{\rm eff}$$=$260 K variation on the T$_{\rm eff}$ corresponding to the 
$\chi$$^{2}_{min}$ solution. \\
In summary, the maximum effect that we measured on the T$_{\rm eff}$ due to rotation is between 
1.5 and 5\% for the more rapidly rotating stars of the sample. This is quite a small effect but 
we stress that its intensity should be checked in any study where the T$_{\rm eff}$ is relevant. 

\begin{table*}[htb] 
\caption[]{T$_{\rm eff}$s derived from the BaSeL library.  
The quoted error is the 1$\sigma$ error.  
The relative error on T$_{\rm eff}$(BaSeL) is given when better than 15\%. 
N is the number of colours to be fitted. Results from spectroscopy (S), MS95 and tB00 are 
also given.} 
\label{tab:finalteff}
\begin{flushleft}
\scriptsize
\begin{center}  
\begin{tabular}{rcrrrrrr}
\hline 
\noalign{\smallskip} 
 ID$^{(\dag)}$ & Comp. & \multicolumn{6}{c}{ ---------------------------------------------------- T$_{\rm eff}$ [K] ---------------------------------------------------- } \\
             &       &  BaSeL & $\sigma$(T$_{\rm eff}$) / T$_{\rm eff}$ & $\chi$$^{2}_{min}$/N & S$^{(1)}$ & MS95 & tB00 \\
\noalign{\smallskip}
\hline \noalign{\smallskip}
1            & p &  5400 $^{+180}_{-140}$    &   3.0 & $<$1   &  5524$\pm$50 &         &    5637 $\pm$ 70 \\
             & s &  4220 $^{+420}_{-340}$    &   9.1 & $<$1   &          &         &    4350 $\pm$ 200 \\
2            & p & $\sim$ 17280              &       & $<$1   &          &         &   11900 $\pm$ 750 \\
             & s & $\sim$ 17060              &       & $<$1   &          &         &   10500 $\pm$ 750 \\
3            & p &  6540 $^{+160}_{-160}$    &   2.4 & $<$1   &  6462    &   6685  &    6890 $\pm$ 100 \\
             & s &  6180 $^{+220}_{-300}$    &   4.3 & $<$1   &          &         &    6280 $\pm$ 70 \\
4            & p &  4920 $^{+150}_{-100}$    &   2.6 & $<$1   &  5100    &         &    5150 $\pm$ 100 \\
             & s &  7820 $^{+580}_{-720}$    &   8.4 & $<$1   &          &         &    7580 $\pm$ 300 \\
5            & p &  $>$ 24500                &       & $<$1   &          &         &    9520 $\pm$ 1200\\
             & s &  7700 $^{+1200}_{-750}$   &  13.0 & $<$1   &          &         &    9230 $\pm$ 2240\\
6            & p &  7000 $^{+180}_{-200}$    &   2.6 & $<$1   &  7128$^{+123}_{-78}$$^{(2)}$   &         &    7200 $\pm$ 150\\
             & s &  5880 $^{+720}_{-660}$    &  11.2 & $<$1   &          &         &    5850 $\pm$ 600\\
7            & p &  5960 $^{+120}_{-140}$    &   2.2 & $<$1   &  5998$\pm$50$^{(3)}$  &         &    6200 $\pm$ 40\\
             & s &  6100 $^{+200}_{-200}$    &   3.3 & 1.1    &          &         &    6280 $\pm$ 40\\
8            & p & $\sim$ 34980              &       & $<$1   & 33000$\pm$5000$^{(4)}$  &         &  18700 $\pm$ 1500 \\
             & s & $>$ 19500$^{(a1)}$        &       & $<$1   &          &         &   22000 $\pm$ 4000\\
9            & p & $\sim$ 15140              &       & $<$1   & 19470$\pm$1830$^{(4)}$  &         &  11900 $\pm$ 625 \\
             & s & $>$ 9980$^{(a2)}$         &       & $<$1   &         &          &   13000 $\pm$ 2800\\
10           & p &  4800 $^{+150}_{-180}$    &   3.5 & $<$1   &   5070  &          &    4900 $\pm$ 275\\
             & s &  6060 $^{+520}_{-560}$    &   8.9 & $<$1   &         &          &    6100 $\pm$ 400\\
11           & p &  6620 $^{+160}_{-220}$    &   2.9 & $<$1   &   6590$\pm$100 &    6547  &    6740 $\pm$ 100\\
             & s &  $\sim$ 5060              &       & $<$1   &         &          &    5250 $\pm$ 400\\
12           & p &  5740 $^{+160}_{-140}$    &   2.6 & $<$1   &   5950$\pm$30  &          &    6220 $\pm$ 43 \\
             & s &  5360 $^{+360}_{-220}$    &   5.6 & $<$1   &   5650$\pm$50  &          &    5410 $\pm$ 110 \\
13           & p &  $\sim$ 9140              &       & $<$1   &   9500$\pm$250  &          &    9520 $\pm$ 317\\
             & s &  $\sim$ 8760              &       & $<$1   &         &          &   11900 $\pm$ 650\\
14           & p &  6360 $^{+600}_{-340}$    &   7.7 & $<$1   &         &   6403   &    6440 $\pm$ 86\\
             & s &  6440 $^{+460}_{-610}$    &   8.4 & $<$1   &         &          &    6378 $\pm$ 140\\
15           & p &  5680 $^{+220}_{-130}$    &   3.2 & 1.8    &         &          &    6280 $\pm$ 60\\
             & s &  4400 $^{+420}_{-200}$    &   7.5 & 1.0    &         &          &    4590 $\pm$ 95\\
16           & p &  6020 $^{+80}_{-70}$      &   1.2 & $<$1   &         &   6093   &    6200 $\pm$ 40\\
             & s &  5920 $^{+140}_{-100}$    &   2.1 & $<$1   &         &          &    6030 $\pm$ 43\\
17           & p & 12100 $^{+1150}_{-600}$   &   7.3 & 3.0    &         &          &   10500 $\pm$ 245\\
             & s &  9160 $^{+390}_{-360}$    &   4.0 & $<$1   &         &          &    8720 $\pm$ 120\\
18           & p & $\sim$ 12820              &       & $<$1   &         &          &    9520 $\pm$ 70\\
             & s & $\sim$ 10740              &       & $<$1   &         &          &    9520 $\pm$ 730\\
19           & p &  9980 $^{+520}_{-280}$    &   4.2 & $<$1   &  8690$\pm$79$^{(5)}$ &          &    9230 $\pm$ 130\\
             & s & 10240 $^{+1060}_{-640}$   &   8.6 & $<$1   &         &          &    9230 $\pm$ 130\\
20           & p &  6780 $^{+520}_{-280}$    &   6.2 & 2.0    &         &    6740  &               \\
             & s &  6120 $^{+610}_{-490}$    &   9.0 & 1.1    &         &          &               \\
21           & p &  9140 $^{+260}_{-240}$    &   2.7 & $<$1   &         &          &    8460 $\pm$ 130\\
             & s &  8940 $^{+310}_{-340}$    &   3.6 & $<$1   &         &          &    8460 $\pm$ 250\\
22           & p & 15800 $^{+1000}_{-1800}$  &   9.2 & $<$1   &         &          &    8460 $\pm$ 250\\
             & s & 12920 $^{+2080}_{-1520}$  &  14.1 & 6.0    &         &          &    8720 $\pm$ 280\\
23           & p &  6280 $^{+120}_{-160}$    &   2.3 & $<$1   & 6240$\pm$270$^{(6)}$ &   6435   &  6440 $\pm$ 60\\
             & s &  6420 $^{+460}_{-400}$    &   6.7 & $<$1   &         &          &    6890 $\pm$ 370\\
24           & p &  8600 $^{+250}_{-220}$    &   2.7 & $<$1   &         &          &    8460 $\pm$ 130\\
             & s &  6440 $^{+860}_{-740}$    &  12.5 & $<$1   &         &          &    6890 $\pm$ 540\\
25           & p &  5900 $^{+100}_{-90}$     &   1.6 & 1.2    & 6390$\pm$150$^{(7)}$ &   6246   &  6360 $\pm$ 40\\
             & s &  6560 $^{+180}_{-180}$    &   2.7 & $<$1   &         &          &    6360 $\pm$ 215\\
26           & p &  6580 $^{+120}_{-130}$    &   1.9 & $<$1   &  6800$^{(8)}$ &   6675   &    6740 $\pm$ 75\\
             & s &  6040 $^{+490}_{-380}$    &   7.1 & $<$1   &         &          &    6280 $\pm$ 210\\
27           & p &  7680 $^{+240}_{-300}$    &   3.5 & 18.7   &         &   6031   &    6030 $\pm$ 840\\
             & s &  5500 $^{+350}_{-600}$    &   8.9 &  6.7   &         &          &    5703 $\pm$ 1090\\
28           & p &  7200 $^{+120}_{-280}$    &   3.0 & $<$1   &         &          &                \\
             & s &  5840 $^{+700}_{-590}$    &  11.1 & $<$1   &         &          &                \\
\noalign{\smallskip}\hline        
\end{tabular}
\end{center}
\scriptsize
(\dag) Arbitrary running number.  
(a$i$) Best solution at 39980 K$^{(a1)}$, 16300 K$^{(a2)}$. 
(1) Cayrel de Strobel {\al} catalogues (1997, 2001), except when mentioned.
(2) Gardiner {\al} (1999) results from H$_{\beta}$, excluding the result with 
overshooting. (2) Perryman {\al} (1998). 
(3) Sokolov (1995), with E(B$-$V)$=$0.07. (4) Blackwell \& Lynas-Gray (1998) by Infrared Flux Method.  
(5) Sokolov (1998). (6) Gardiner {\al} (1999). (7) Upper limit from the H$_{\beta}$ line (Solano \& Fernley, 1997). 
\normalsize
\end{flushleft}
\end{table*}

\section{Results and discussion}
\label{section:results}
Our final effective temperature results are listed in Tab.~\ref{tab:finalteff}, along 
with earlier determinations. A direct comparison of these determinations is discussed in 
more detail in \S\ref{section:comparison1}. 
Useful information is given in Tab.~\ref{tab:finalteff} for a better interpretation 
of the T$_{\rm eff}$ determinations from the BaSeL models: the relative error on the 
T$_{\rm eff}$ when better than 15\%, 
and the $\chi$$^2$ defined in Eq. 1 normalized to the number of fitted colours. 
For clarity of Tab.~\ref{tab:finalteff}, 
all good fits are simply labelled by a $\chi^2$ less than 1. 
The photometric accuracy being often better for the primary stars (see Tab. 5 of tB00), 
the contours and hence the T$_{\rm eff}$ determinations are less strongly 
constrained for the secondary components.  
Finally, it has to be mentioned that the systems [21] and [22] are known to have 
at least a third component ({\eg} [22] is a multiple system that contains at least 
5 stars, Mc Kibben {\al} 1998).   
Since these components are not resolved in the photometric 
measurements of ten Brummelaar {\al} (2000), the derived temperatures might be  
less reliable than for the other stars listed in Tab.~\ref{tab:finalteff}. 
 
\subsection{General comparison with earlier results}
\label{section:comparison1}
\begin{figure*}[htb]
\centerline{\psfig{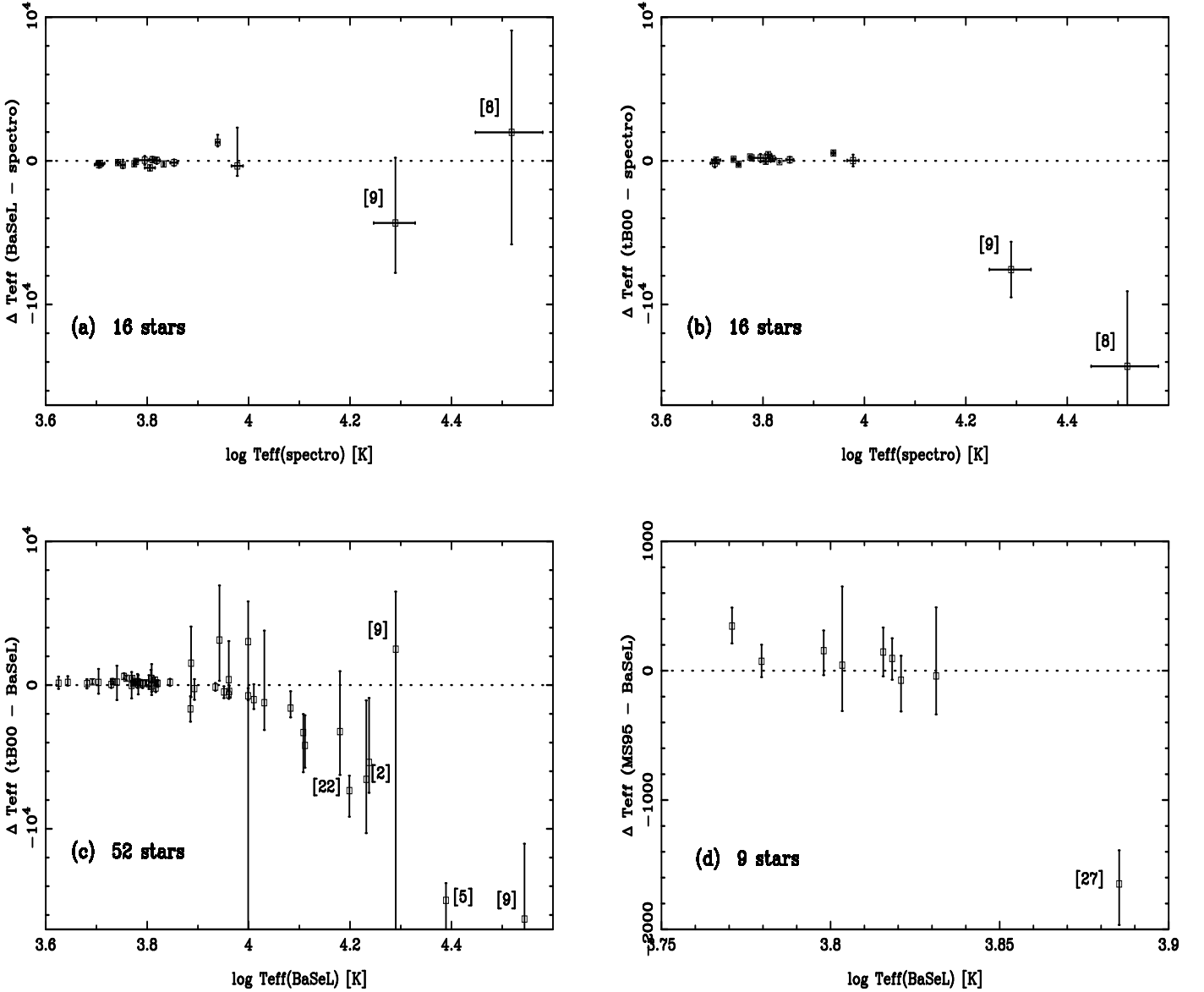}}
\caption{
Comparisons of T$_{\rm eff}$ determinations: (a) BaSeL and spectroscopic determinations,  
(b) tB00 and spectroscopic determinations, (c) tB00 and BaSeL determinations, and (d) 
MS95 and BaSeL determinations. 
For eye-guidance, a line of equal temperature (dotted line) is shown on each panel. 
}
\label{f:teff_comp}
\end{figure*}
Fig.~\ref{f:teff_comp} compares various T$_{\rm eff}$ determinations: Marsakov \& Shevelev (1995), 
ten Brummelaar {\al} (2000), spectroscopic determinations and the present work with the BaSeL 
library. Both tB00 and BaSeL results show equally good agreement with 
spectroscopic determinations for temperatures cooler than $\sim$10,000 K (see panels (a) and (b)). 
The agreement with spectroscopy is better with BaSeL than tB00 results beyond 10,000 K, 
even if this conclusion comes from only two objects (primaries of systems [8] and [9]). 
A direct comparison between tB00 and BaSeL solutions (panel (c)) confirms the previous 
deviation: 
the agreement is good for temperatures cooler than $\sim$10,000 K, but tB00 temperatures are 
systematically and increasingly cooler for hotter temperatures.  
Another comparison is shown in panel (d) between BaSeL and MS95 determinations 
(derived from Str\"omgren photometry): the stars in common show a good agreement in the range 
6000-6700 K. The only disagreement appears for the primary of the system [27] but in this 
case we know that BaSeL is not able to fit simultaneously the three colour indices B$-$V, 
V$-$R and V$-$I (see bad $\chi$$^2$ in Tab.~\ref{tab:finalteff}). 
As a matter of fact, $\chi$$^2$-experiments with the BaSeL models show that a good fit is 
obtained for this star if only B$-$V and V$-$R are kept (hence excluding V$-$I data). 
In this case, we would derive T$_{\rm eff}$$=$ 7940 K (assuming E(B$-$V)$=$ 0) and 8000 K 
(assuming E(B$-$V)$=$ 0.006). 
In both cases the disagreement shown in panel (d) between MS95 and BaSeL would 
remain.  \\
\begin{figure}[htb]
\centerline{\psfig{file=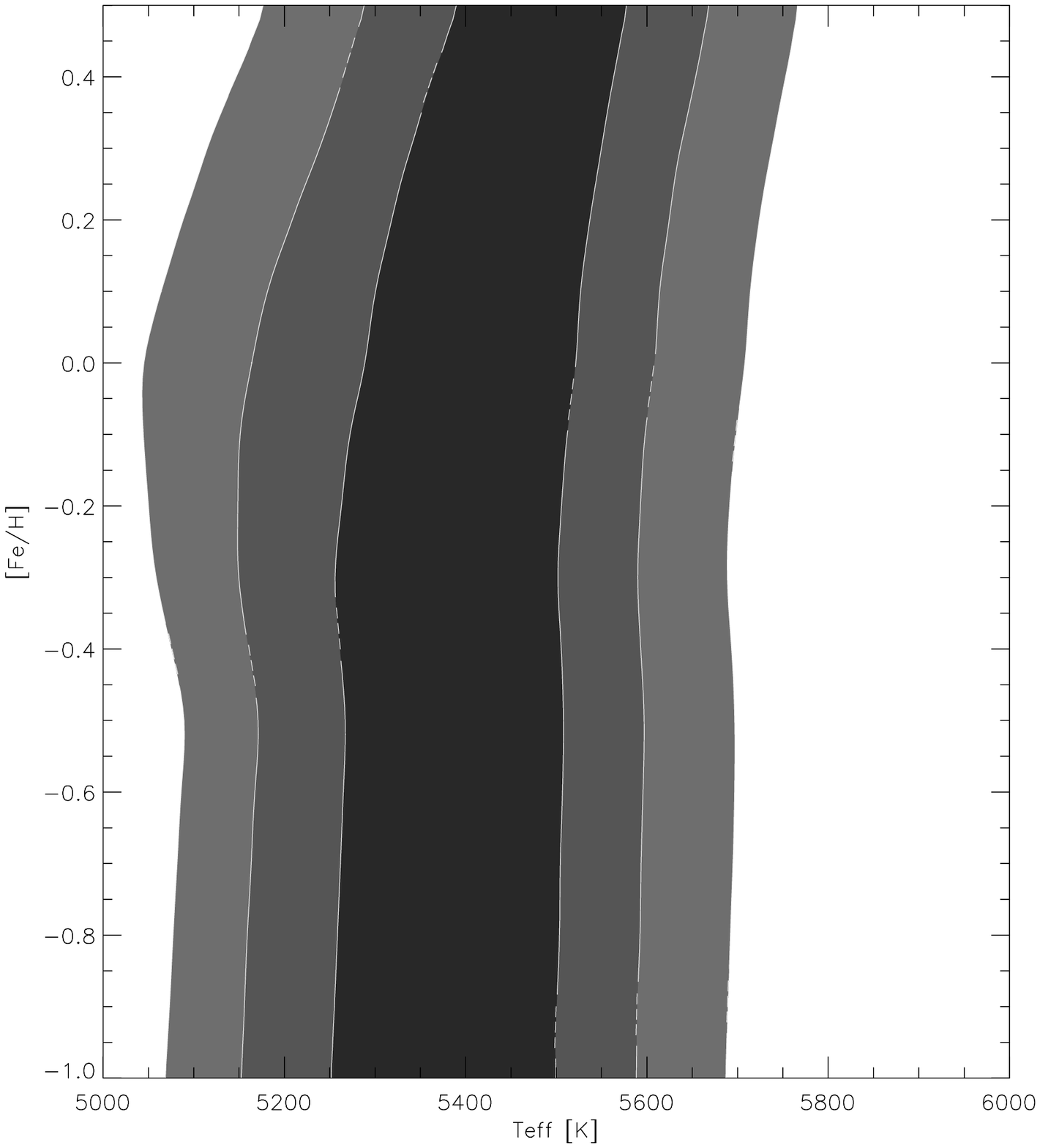,width=9.truecm,height=7.5truecm,rheight=7.9truecm}}
\caption{Contour solutions for the primary of 85 Peg ([1]) in the (T$_{\rm eff}$, [Fe/H]) 
plane by fitting the dereddened observed (B$-$V), (V$-$R) and (V$-$I) colours with the BaSeL 
library.   
}
\label{f:85PegA}
\end{figure}
\begin{figure*}[htb]
\centerline{
\psfig{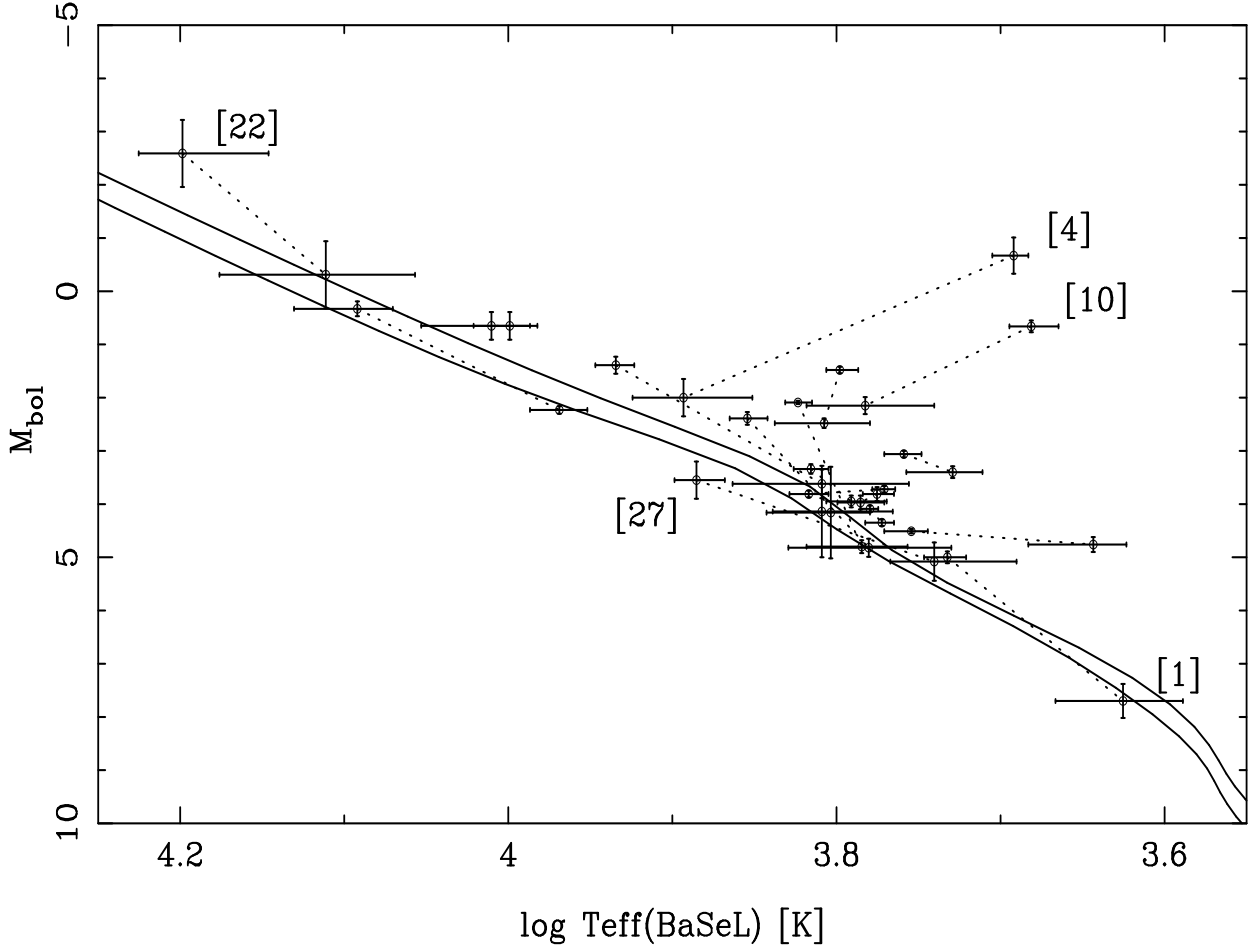}}
\caption{
HR diagram with BaSeL T$_{\rm eff}$ determinations for the systems with 
a relative accuracy on their effective temperature better than 15\% for both components. 
The components of each system are joined by a dotted line. The correction for rotation 
has been applied according to Tab.~\ref{tab:rotation}.
The system [22] is labelled because its T$_{\rm eff}$s might be less reliable 
as explained in \S\ref{section:results}.
The system [27] is also shown because both components present 
a bad fit to the observed colours (see Tab.~\ref{tab:finalteff}).
The systems [4] and [10] contain an evolved star (see text). 
Analytical ZAMS sequences from Tout {\al} (1996) are displayed for eye-guidance 
for solar metallicity ([Fe/H]$=$0) and the metallicity of 85 Peg [1] ([Fe/H]$=$$-$0.57).  
}
\label{f:hrd}
\end{figure*}
\subsection{More detailed comparisons for two selected binaries}
\label{section:comparison2}
Among the working sample, one of the most studied binary system is 85 Peg ([1] in Tab.~\ref{tab:data}). 
According to spectroscopy, its primary component has an effective temperature of 5524$\pm$50 K 
(Axer {\al} 1994). This confirms the quality of the photometric determinations given in 
Tab.~\ref{tab:finalteff} 5400$^{+180}_{-140}$ K (BaSeL) and 5637$\pm$70 K (tB00). 
An advantage of the BaSeL results is that the T$_{\rm eff}$ is metallicity dependent, as 
depicted by Fig.~\ref{f:85PegA}. Therefore, assuming a value of [Fe/H] induces a 
different T$_{\rm eff}$ solution due to the iso-contours shape.      
If one adopts the most recent estimation of [Fe/H] ($-$0.57$\pm$0.11, Fernandes {\al} 2002), 
the corresponding T$_{\rm eff}$ solutions are sligthly reduced and cover the 5270-5510 K range   
whose upper limit is consistent with the spectroscopy. 
On the other hand, if the T$_{\rm eff}$ is fixed to its spectrocopic value, the contour of 
Fig.~\ref{f:85PegA} would suggest a [Fe/H] $\geq$ $-$0.3. \\
There is no spectroscopic determination for the secondary component of 85 Peg 
for which we infer T$_{\rm eff}$$=$4220 $^{+420}_{-340}$ K from the BaSeL models. 
This result is in good agreement with tB00 
(see Tab.~\ref{tab:finalteff}). \\
Another system - $\xi$ UMa ([12] in Tab.~\ref{tab:data}) - is also of particular interest because 
it is the only one in Tab.~\ref{tab:data} with both components having a spectroscopic determination of 
their fundamental stellar parameters: (T$_{\rm eff}$, [Fe/H]) $=$ (5950 K, $-$0.35) (primary) 
and (5650 K, $-$0.34) (secondary). 
The temperatures derived by tB00 are either larger (by more than 6$\sigma$ for 
HD 98231, the primary component) or smaller (by more than 2$\sigma$ for HD 98230, the 
secondary component) than the spectroscopic values. 
Our determinations with BaSeL are smaller than the spectroscopic values for both components 
but only inside the 1$\sigma$ uncertainty for the secondary, and at $\sim$1.3$\sigma$ for 
the primary component. However, this better agreement is partly due to the large error 
bars. 
Since the BaSeL results are metallicity-dependent, the results reported 
in Tab.~\ref{tab:finalteff} are given for [Fe/H] in the range [$-$1, 0.5]. 
If one assumes the spectroscopic values of the iron abundance, we derive temperature 
solutions slightly greater and with smallest uncertainties: 
T$_{\rm eff}$($\xi$ UMa A)$=$ 5765$\pm$85 K and 
T$_{\rm eff}$($\xi$ UMa B)$=$ 5435$\pm$215 K, 
in a better agreement with the spectroscopic values.  

\subsection{The revised HR diagram}
Fig.~\ref{f:hrd} shows a M$_{\rm bol}$-log T$_{\rm eff}$ diagram, revision of the HR diagram 
presented by ten Brummelaar {\al} (2000). 
A subsample of the systems with $\sigma$(T$_{\rm eff}$)/T$_{\rm eff}$$\leq$0.15 for both 
components (see Tab.~\ref{tab:finalteff}) is shown, 
excluding 7 systems from the working sample: [2], [5], [8], [9], [11], [13] and [18].  
This selection on the T$_{\rm eff}$ relative accuracy covers a large range of temperatures  
(from $\sim$4000 to 16000 K) and, hence, is of particular interest for future tests of 
stellar evolution models, in particular when the metallicity of the system is 
spectroscopically known. 
For this reason we provide in Tab.~\ref{tab:bestteff} the T$_{\rm eff}$ fixing  
[Fe/H] to its spectroscopic value\footnote{We exclude the system [13] from 
Tab.~\ref{tab:bestteff} because the uncertainty on T$_{\rm eff}$ remains 
larger than 15\% even fixing [Fe/H] to its spectroscopically value ($-$0.02).}
: the ranges are substantially reduced in comparison 
to the results listed in Tab.~\ref{tab:finalteff} where all the [Fe/H] values from 
$-$0.50 to 1.00 were considered. 
This illustrates a direct use of the metallicity-dependent T$_{\rm eff}$ derived from 
the BaSeL iso-contours. 
This subsample with spectroscopic [Fe/H] determinations and revised T$_{\rm eff}$s for both 
components should be considered in priority for further applications. In particular,  
the systems [4] and [10] have one main sequence star and one evolved component: this  
is useful for testing isochrones on different evolutionary phases. 
The binaries [6] and [7] belongs to the Hyades open cluster.
The other systems listed in Tab.~\ref{tab:bestteff} are close to the Zero Age Main Sequence 
(ZAMS) and are therefore promising to check ZAMS models for metallicities ranging from 
$-$0.57 to 0.16.  
  
\begin{table}[htb] 
\caption[]{T$_{\rm eff}$ results derived from the BaSeL library assuming the spectroscopic 
value of [Fe/H] from Cayrel de Strobel {\al} catalogues, except for [1] 
(Fernandes {\al}, 2002) and [23] (Sokolov, 1998).}  
\label{tab:bestteff}
\begin{flushleft}
\begin{center}  
\begin{tabular}{rrrr}
\hline 
\noalign{\smallskip} 
 ID$^{(\dag)}$ & [Fe/H] & \multicolumn{2}{c}{  T$_{\rm eff}$ [K]  } \\
               &        &   primary & secondary  \\
\noalign{\smallskip}
\hline \noalign{\smallskip}
1            & $-$0.57 &  5390 $\pm$ 120        &   4265 $\pm$ 335   \\
3            & $-$0.26 &  6525 $\pm$ 125        &   6170 $\pm$ 190   \\
4            & $-$0.17 &  4870 $\pm$ 50         &   7650 $\pm$ 450   \\
6            &    0.14 &  6960 $\pm$ 140        &   5885 $\pm$ 665   \\
7            &    0.16 &  5935 $\pm$ 105        &   6100 $\pm$ 190   \\
10           &    0.05 &  4765 $\pm$ 65         &   6020 $\pm$ 500   \\
11           & $-$0.30 &  6565 $\pm$ 115        &   5125 $\pm$ 675   \\
12           & $-$0.35 &  5765 $\pm$ 85         &   5435 $\pm$ 215   \\
23           &    0.00 &  solution at 3$\sigma$ &  6330 $\pm$ 280    \\
\noalign{\smallskip}\hline        
\end{tabular}
\end{center}
(\dag) Arbitrary running number.  
\end{flushleft}
\end{table}

\section{Conclusion}
The agreement that we obtain between the Alonso {\al} (1996) empirical 
calibrations and BaSeL 2.2 for dwarf stars in the range 4000-8000 K  
fully justifies to determine the effective temperature from the Johnson  
photometry of the theoretical BaSeL library. 
In this context, we have presented new homogeneous T$_{\rm eff}$ determinations for 
each component of a sample of 28 binary stars from BaSeL synthetic photometry. 
As expected from BVRI colour combinations, we did not obtain useful constraints 
on the surface gravity and the metallicity because these colours are not very 
sensitive to these parameters. 
Nevertheless our solutions give metallicity-dependent T$_{\rm eff}$s, which 
is of particular interest when [Fe/H] is known. 
This sample is of particular importance because there are relatively few systems 
for which both individual components can be placed in a HRD diagram,  
except some eclipsing binaries (see {\eg} Lastennet {\al} 1999a, b) 
or nearby visual binaries ({\eg} Fernandes {\al} 1998, Morel {\al} 2001). 
For this reason, we paid particular attention to the influence of reddening and 
stellar rotation before deriving their T$_{\rm eff}$. 
We derived the reddening from two different methods: (i) the MExcessNg code v1.1 
(M\'endez \& van Altena 1998) and (ii) neutral hydrogen column density data. 
A comparison of both methods shows a good agreement for our small sample, but 
we point out a few directions where the MExcess model overestimates the 
E(B$-$V) colour excess. 
As for the influence of stellar rotation on the BVRI colours, we neglected it 
except for the 5 stars with large $v \sin i$. However, even in these cases the shift 
in temperature is about 5\% at maximum. 
Our final determinations provide effective temperature estimates for
each component (see Tab. \ref{tab:finalteff}). 
They are in good agreement with previous spectroscopic 
determinations available for a few primary components, and suggest that earlier 
determinations from tB00 are systematically and increasingly underestimated 
beyond 10,000 K. A revised HR diagram is provided, with a selection of binaries 
with relative accuracy on the T$_{\rm eff}$ better than 15\%. Finally, 
reduced uncertainties on the T$_{\rm eff}$ determinations are given by fixing 
[Fe/H] to its spectroscopically value when available. This subsample should 
be considered in priority for further applications ({\eg} calibration of 
stellar evolution models), in particular the systems 
[4] and [10] because they have now 1) accurate T$_{\rm eff}$ determinations 
for both components 2) [Fe/H] determination from spectroscopy and 3) one 
main sequence star and one evolved component which is useful for testing isochrones  
on different evolutionary phases.  

\begin{acknowledgements}
EL thanks Ren\'e A. M\'endez for kindly providing his reddening model source code, 
and Marian Douspis for his IDL computing skills. It is a pleasure to thank 
Edouard Oblak for useful suggestions about the colour excess aspect of this paper. 
We thank the referee, T. ten Brummelaar, for his careful reading of the manuscript. 
EL and TL are supported by the "Funda{\c{c}}\~ao para a  Ci\^encia e Tecnologia'' 
(FCT) postdoctoral fellowships (grants SFRH/BPD/5556/2001 and 
PRAXIS-XXI$/$BPD$/22061/99$ respectively). This work was partially 
supported by the project "PESO/P/PRO/15128/1999" from the FCT. 
This research has made use of the SIMBAD database operated at CDS, Strasbourg, France, 
and of NASA's Astrophysics Data System Abstract Service. 
\end{acknowledgements}


\end{document}